\documentclass{optica-article}

\journal{opticajournal} % for journals or Optica Open

\articletype{Research Article}

\begin{document}

\title{Suppression of the Talbot effect in Fourier Transform Acousto-Optic imaging}

\author{Maïmouna Bocoum\authormark{1*}, François Figliolia\authormark{1}, Jean-Pierre Huignard\authormark{1}, François Ramaz\authormark{1}  and Jean-Michel Tualle\authormark{2}}

\address{\authormark{1}Institut Langevin, ESPCI Paris, Université PSL, CNRS, 75005 Paris, France\\
\authormark{2}Laboratoire de Physique des Lasers, CNRS UMR 7538, Université Sorbonne Paris Nord, 99 avenue J.-B.~Clément, 93430 Villetaneuse, France\\}

\email{\authormark{*}maimouna.bocoum@espci.fr} %% email address is required; see note below about the corresponding author designation

\begin{abstract*} 
We report on the observation and correction of an imaging artifact attributed to the Talbot effect in the context of acousto-optic imaging using structured acoustic waves. When ultrasound waves are emitted with a periodic structure, the Talbot effect produces $\pi$-phase shifts of that periodic structure at every half of the Talbot distance in propagation. This unwanted artifact is detrimental to the image reconstruction which assumes nearfield diffraction is negligible. Here, we demonstrate both theoretically and experimentally how imposing an additional phase modulation on the acoustic periodic structure induces a symmetry constraint leading to the annihilation of the Talbot effect. This will significantly improve the acousto-optic image reconstruction quality and allows for an improvement of the reachable spatial resolution of the image. 
\end{abstract*}

%%%%%%%%%%%%%%%%%%%%%%%%%%  body  %%%%%%%%%%%%%%%%%%%%%%%%%%

\section{Introduction}

Acousto-Optic (AO) is an in-depth optical imaging technique developed for highly scattering media for which conventional imaging is challenging. It involves the use of controlled ultrasonic waves (US) to tag photons, and a reconstruction method that is highly dependent upon the choice of US spatio-temporal profile. The use of periodically structured insonification to perform AO imaging has been demonstrated~\cite{bocoum2019structured,bocoum2020reconstruction,dutheil2021fourier}. In particular, the Fourier Transform Acousto-Optic imaging (FT-AOI) method~\cite{bocoum2020reconstruction,dutheil2021fourier} is well adapted to digital holographic-based detection of tagged photons. That latter detection is compatible with \textit{in-vivo} AO imaging provided the exposure time of the camera be less than the medium decorrelation time, typically $\lesssim~\mathrm{ms}$ in biological tissues~\cite{qureshi2017vivo,liu2015optical,cheng2023high}. In FT-AOI, a monochromatic US plane wave is modulated in amplitude along both the direction of propagation and the direction along the emission probe. A similar amplitude modulation is applied to the reference beam used as the Local Oscillator (LO) on the camera, such that detected tagged photons are periodically located in the US imaging plane. This feature gives access to a Fourier component of the image to reconstruct. Using several structuring harmonics components and phase offsets, the whole complex Fourier plane of the image can be fetched. The image is then simply reconstructed by inverse Fourier transform. 
In previous work~\cite{dutheil2021fourier}, the spatial resolution of the images recorded using FT-AOI remained however moderate, up to about $8\lambda \sim 4~\mathrm{mm}$, \textit{i.e.} almost an order of magnitude away from diffraction limited resolution. In our previous attempt to improve the spatial resolution by increasing the structuration frequency along the transverse direction of the transducer, we systematically observed a strong degradation of the image. In this article, we explain the origin of this degradation we attribute to the Talbot effect. We will see how imposing an additional phase modulation on the periodic US pulse allows to get rid of the artifact and reach near diffraction-limited imaging resolutions.\\

\newpage

The Talbot effect is a near field diffraction phenomenon discovered by H.F. Talbot in 1836~\cite{talbot1836lxxvi} who observed that a grating illuminated by a spatially coherent source would produce images of itself close to its surface as a result of free space propagation. These so called revival images are periodically separated by the Talbot distance:

\begin{equation}
z_T = \frac{2a^2}{\lambda},
\end{equation}

\noindent with $a$ is the typical period of the grating, and $\lambda$ the optical wavelength. This effect, which generalizes to any spatially periodic coherent source, has found multiple applications in wave physics. A non-exhaustive list of these applications include the use of laser Talbot cavities to phase lock semiconductor laser arrays~\cite{leger1989lateral,leger1988coherent,mehuys1991modal}, structured illumination in fluorescence microscopy~\cite{chowdhury2018structured} or lithography~\cite{stuerzebecher2010advanced,wang2016sub}. 
Although the Talbot effect was first observed and studied in optics~\cite{rayleigh1881xxv} within the scope of Fresnel diffraction theory, it has also been studied for other types of wave physics such as plasmonics~\cite{dennis2007plasmon}, matter waves~\cite{chapman1995near,clauser1994talbot,deng1999temporal}, electromagnetic or more recently ultrasonic waves~\cite{morozov2017observation,candelas2019observation}. One interesting feature of the Talbot effect  is the concomitant appearing of self-images shifted by half a period also appear at a fraction of $z_T$ on what is designated as the ``secondary Talbot image'', as well as more complicated image revival at fractional multiples of the Talbot distance~\cite{latimer1992talbot,berry2001quantum}. Although the richness of the Talbot carpet might be used for quantum revivals~\cite{wen2013talbot} or the investigation of number theory~\cite{pelka2018prime,de2013theory}, it turns out to be a detrimental artifact to any imaging method where near field diffraction effects are neglected, as is the case in FT-AOI. It has recently been shown how the Talbot effect can be exploited to shape frequency combs using phase control in the temporal or spectral domain~\cite{romero2019arbitrary}. In this article, we resort to this method and show how imposing an additional $\pi$-phase modulation to the periodic pattern allows to remove this artifact and greatly improve the resolution of the imaging technique.

\section{Theoretical description of FT-AOI}

\begin{figure}[htbp]
\centering\includegraphics[width=\textwidth]{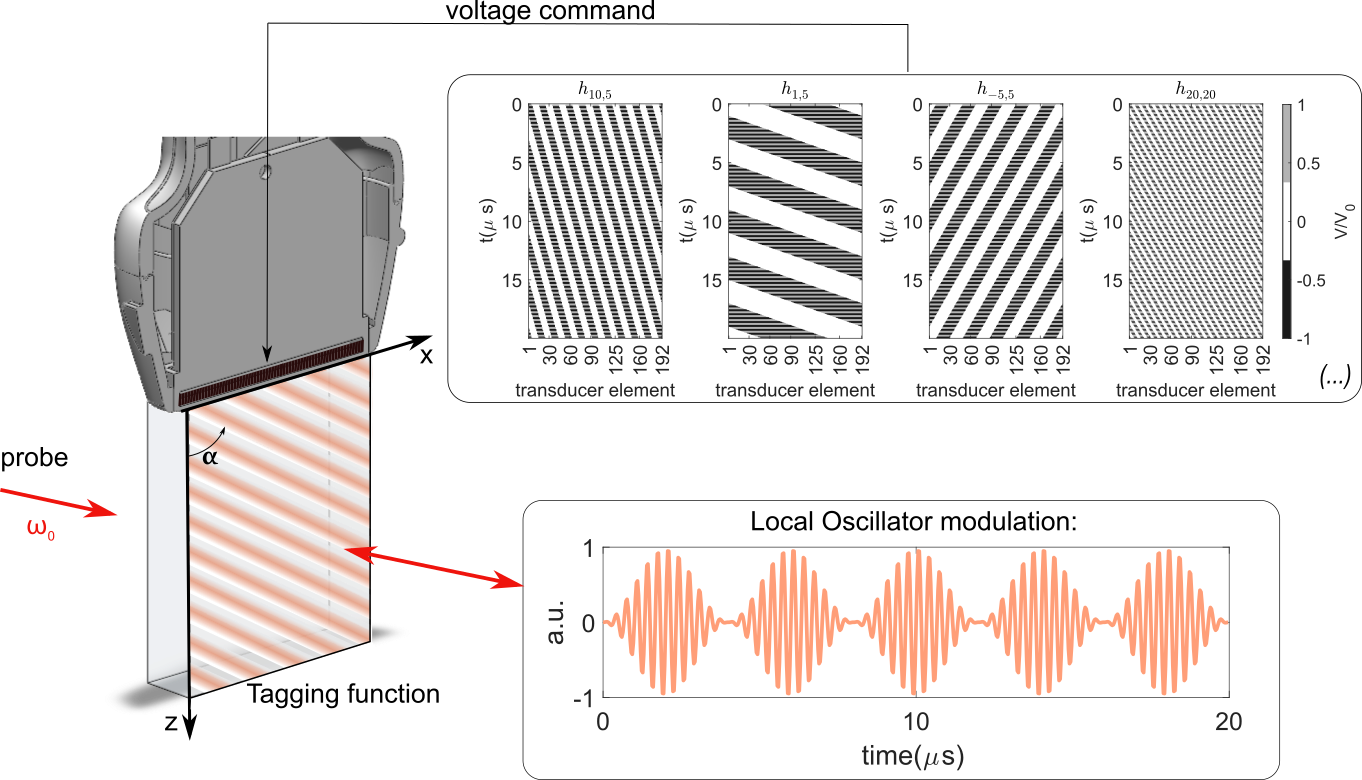}
\caption{Illustration of the different periodic amplitude modulations $h_{n,m}$ imposed on the command voltage $V$ applied the transducer (top caption), where $V_0$ designates a nominal voltage set by the user. The electronics of the transducer only allows for $[-V_0 , 0 , V_0]$ as driving voltage, but the resulting emitted pressure field is smoothed out by the limited spectral bandwidth of the transducer array.  Artistic representation of an ideal tagging function for $h_{n=10,m=5}$ resulting from off-axis holographic detection using the local oscillator which phase modulation is depicted the bottom right for $n=5$ and $\phi=0$.}
\label{fig:scheme}
\end{figure}

Let's begin the description of the method as described in~\cite{dutheil2021fourier}, \textit{i.e.} neglecting diffraction of the propagating US. In FT-AOI, photons propagating in a scattering media are tagged by means of a long US pulse ($>100~\mathrm{\mu s}$ typically, carrier in the MHz range) periodically modulated in amplitude both along the propagation direction $z$, and the US transducer transverse direction $x$. The command voltage applied on a transducer array (length $L$, composed of 192 elements) is illustrated in the top caption Fig~\ref{fig:scheme}, where different structurations are depicted.

\noindent The generated pressure field $P_{n,m}$ corresponding to the modulation with harmonics $(n,m)$, where $n>0$ and $m$ are integers, can be expressed in complex form:

\begin{equation}
\label{eq:hyp}
P_{n,m}(x,z,t) = P_0 h_{n,m}(x,z,t)e^{2i \pi \nu_{us}(t-c_s^{-1}z)},
\end{equation}

\noindent where $x,z$ are the coordinates of a point in the imaging plane, $t$ is the time following the emission by the transducer array,$c_s$ is the sound velocity, $P_0$ is the nominal pressure field directly set by the driving voltage $V_0$, $\nu_{us}$ the ultrasonic carrier frequency, and where $h_{n,m}(x,z,t)$ is the amplitude modulation function. The command voltage applied to the transducer is a $T_0/n$-periodic function $h_n(\cdot)$ with a phase shift proportional to the position $x$, with $T_0$ a fundamental period freely set by the user. We typically set it such that $ T_0$ be the same order of magnitude as $c_s^{-1}L$. If one neglects diffraction effects, the modulation function is simply obtained from $h_n$ using the relation:

\begin{equation}
\label{eq:Eq2bis}
h_{n,m}(x,z,t)\equiv h_n( t - \alpha x - c_s^{-1}z),
\end{equation} 

\noindent with $\alpha=mT_0/(nL)$ the appearing angle of the modulation illustrated in Fig~\ref{fig:scheme}. We perform digital off-axis holographic detection with a local oscillator (LO) centered at $\nu_{0}+\nu_{us}$, where $\nu_{0}$ is the illumination laser carrier frequency, and modulated in amplitude by a sin-wave function $h^{n,\phi}_{\mathrm{ref}}(.)$ with $T_0/n$-periodicity, where $\phi$ a user defined phase-shift. A typical temporal profile of the LO phase modulation is illustrated at the bottom right of Fig~\ref{fig:scheme} for $n=5$ and $\phi=0$. Note that although other LO modulation with $T_0/n$-periodicity could be used, the choice of sin-wave is very well-suited for the detection when accounting for the impulse response of the transducer. The weight contribution of tagged photons issued from position $(x,z)$ to the detected signal will depend on their relative phase offset with the LO. This so called "tagging function" is consequently bi-periodic in the US imaging plane as illustrated by the artistic representation in Fig~\ref{fig:scheme}, with a maximum contribution when both are in phase. To fetch the Fourier component of the object with frequency $(m\nu_{x0},n\nu_{z0})$, where the fundamental frequencies are defined as $\nu_{x0} = 1/L $ and $\nu_{z0} = 1/(c_sT_0) $, we simply need to acquire four consecutive frames with respectively $\phi={0;\frac{\pi}{2};\pi;\frac{3\pi}{2}}$, from which we can estimate a complex Fourier component~\cite{dutheil2021fourier}. By repeating the process for different harmonic orders $(n,m)$, an image is reconstructed by inverse Fourier transform. In previous work~\cite{dutheil2021fourier}, an experimental reconstruction was successfully performed for $-5\le m \le 5$ and $1\le n \le 10$ with $\nu_{x0} = 26.04~\mathrm{m^{-1}}$ and $\nu_{z0} = 32.46~\mathrm{m^{-1}}$, corresponding to a spatial resolution of $\sim~4.6~\mathrm{mm}$ along $x$ and $\sim~1.8~\mathrm{mm}$ along $z$. In order to improve the spatial resolution of the image along x, it is necessary to increase $m$. Doing so however, near-field diffraction of the field known as the Talbot effect will start to manifest (typically for $m\ge 5$), such that any attempt to increase the spatial resolution along $x$ will induce a simultaneous degradation of the reconstructed image. 

\section{Talbot effect in FT-AOI}

\noindent To evaluate the Talbot Effect on the tagging function, we now account for diffraction in the near field of the transducer array for a given value of $(n,m)$. The function $h_{n}(t)$ is $T_0/n$-periodic and can therefore be expressed as a  Fourier series:

\begin{equation}
\label{eq:h_nm}
h_{n}(t) = \sum_{k=-\infty}^{\infty}a^{n}_{k} e^{2i \pi\frac{kn}{T_0}t},
\end{equation}

\noindent where the $a^{n}_{k}$ are the coefficients of the series decomposition. Here, we impose $h_{n}(t)$ to be symmetric with respect to the origin so that $a^{n}_{k}$ are real numbers, without loss of generality. The expression of the US field in the imaging plane $(x,z)$ following the propagation can be derived using the Fresnel propagator:

\begin{equation}
P^{n,m}(x,z,t) = \int_{\mathbb{R}} \left( \int_{\mathbb{R}}  P^{n,m}_0(\nu_x,\nu) e^{2i\pi\frac{\nu_x^2 z c_s}{2\nu}} e^{-i2\pi \nu_x x}d\nu_x \right)e^{i2\pi \nu (t-c_s^{-1}z)}d\nu,
\label{fresnel}
\end{equation}

%\begin{equation}
%P(\g{r},t) = \int_{\mathbb{R}} \left( \iint_{\mathbb{R}^2}  P_0(\nu_x,\nu_y,\nu) e^{j2\pi\frac{(\nu_x^2 + \nu_y^2) z c_s}{2\nu}} e^{-j2\pi(\nu_x x + \nu_y y)}d\nu_xd\nu_y \right)e^{j 2\pi \nu (t-c_s^{-1}z)}d\nu
%\end{equation}

\noindent where $P^0_{n,m}(\nu_x,\nu)$ is the initial spatiotemporal spectrum of the emitted pressure field resulting from the amplitude modulation $h_{n,m}$, defined at $z=0$ in Eq~\ref{eq:Eq2bis}, $\nu$ designates the temporal frequency of the acoustic wave, $\nu_x$ the acoustic spatial frequency.  In this expression, the diffraction effects are embedded in the quadratic phase function $\exp({j\pi\nu_x^2 z c_s/\nu})$. From Eq~\ref{eq:Eq2bis} and \ref{eq:h_nm} we have:
%
%\begin{equation}
%\label{eq:hypt=0}
%P_{n,m}(x,0,t) = P_0\int_{\mathbb{R}} \int_{\mathbb{R}} \sum_{k=-\infty}^{\infty}\sum_{l=-\infty}^{\infty}a^{n,m}_{k,l}\delta(\nu_x - \frac{lm}{L}) \delta(\nu - (\frac{kn}{T_0}+\nu_{us})) e^{-2 j\pi\nu_x x}e^{2j \pi \nu t}  d\nu_xd\nu
%\end{equation}

\begin{equation}
\label{eq:hypt=0}
P_0^{n,m}(\nu_x,\nu) = P_0\sum_{k=-\infty}^{\infty}a^{n}_{k}\delta(\nu_x - \frac{km}{L}) \delta(\nu - (\frac{kn}{T_0}+\nu_{us})) .
\end{equation}

\noindent We inject $P_0^{n,m}(\nu_x,\nu)$ into Eq~\ref{fresnel} which leads to:

\begin{equation}
P^{n,m}(x,z,t) = P_0\sum_{k=-\infty}^{\infty}a^{n}_{k} e^{i\pi (\frac{km}{L})^2\frac{  z c_s}{(\nu_{us}+\frac{kn}{T_0})}} e^{-2 \pi i \frac{km}{L}x}
e^{i 2\pi(\nu_{us} + \frac{kn}{T_0})(t-c_s^{-1}z)}.
\end{equation}

\noindent We can neglect the structuring frequencies relative to the US carrier such that $\nu_{us}\gg \frac{kn}{T_0}$. As a result, we express the modulation function defined in Eq~\ref{eq:hyp} as:

%\begin{equation}
%P(x,z,t) = \int_{\mathbb{R}} \left( \int_{\mathbb{R}} P_0\sum_{k=-\infty}^{\infty}\sum_{l=-\infty}^{\infty}a^{n,m}_{k,l}\delta(\nu_x - \frac{lm}{L}) \delta(\nu - (\frac{kn}{T_0}+\nu_{us})) e^{j2\pi\frac{\nu_x^2 z c_s}{2\nu}} e^{-j2\pi \nu_x x}d\nu_x \right)e^{j2\pi \nu (t-c_s^{-1}z)}d\nu
%\label{fresnel-injected}
%\end{equation}

\begin{equation}
\label{eq:diffHnm}
h^{n,m}(x,z,t) = \sum_{k=-\infty}^{\infty}\left(a^{n}_{k} e^{i\theta_m k^2}e^{-2 i\pi k\chi_{nm}}\right) e^{2i\pi\frac{kn}{T_0}t},
\end{equation}

\noindent with $\theta_m\equiv \pi\frac{m^2}{L^2}\frac{zc_s}{\nu_{us}}$ and $\chi_{nm}\equiv \frac{mx}{L}+\frac{nz}{c_sT_0}$.  
We now define the tagging function $C_{n,m}(x,z)$ resulting from the off-axis holographic measurement by the correlation function:

\begin{equation}
\label{eq:DefCorrel}
C_{n,m}(x,z) = \left|\frac{1}{\tau_e}\int_{\tau_e} h^{n,m}(x,z,t) h^{n,\phi}_{\mathrm{ref}}(t)^*dt \right|^2,
\end{equation}

\noindent where $h^{n,\phi}_{\mathrm{ref}}$ is the amplitude modulation function of the LO~\cite{dutheil2021fourier} and $\tau_e$ the integration time of the camera, which we impose to be a multiple of $T_0$. This correlation function is essential as it represents the spatial position of the tagged photons in the insonification plane. In absence of diffraction artifacts, $C_{n,m}(x,z)$ is a purely periodic function along both $x$ and $z$~\cite{dutheil2021fourier}. We assume a sine-wave for the reference modulation:

\begin{equation}
\label{eq:Href} 
h^{n,\phi}_{\mathrm{ref}}(t) = 1 +\cos\left(2\pi\frac{n}{T_0}t+\phi\right)
\end{equation}

\noindent The correlation in Eq~\ref{eq:DefCorrel} acts as a filter for the expression Eq~\ref{eq:diffHnm}. Using Eq~\ref{eq:Href} and Eq~\ref{eq:diffHnm}, we calculate the correlation function and find the following expression:

\begin{equation}
C_{n,m}(x,z) = |a_0^n + a_1^ne^{i\theta_m}\cos(2\pi\chi_{nm}+\phi)|^2
\label{eq:CnotCorr}
\end{equation}

\noindent This simple expression allows a quite straightforward interpretation of the Talbot effect which manifests on the tagging function: it corresponds to the beating interference between the constant term $a_0^n$ and the bi-periodic structure $a_1^n\cos(2\pi\chi_{nm}+\phi)$ driven by the complex coefficient $e^{i\theta_m}$. As such, a $\pi$-offset on the fringes is observed for $\theta_m = p\pi$ with $p\in \mathbb{Z}$, which corresponds to the positions:

\begin{equation}
 z_p = p \frac{\nu_{us}L^2}{c_sm^2} = p \frac{z_m}{2},
\end{equation}

\begin{figure}[htbp]
\centering\includegraphics[width=\textwidth]{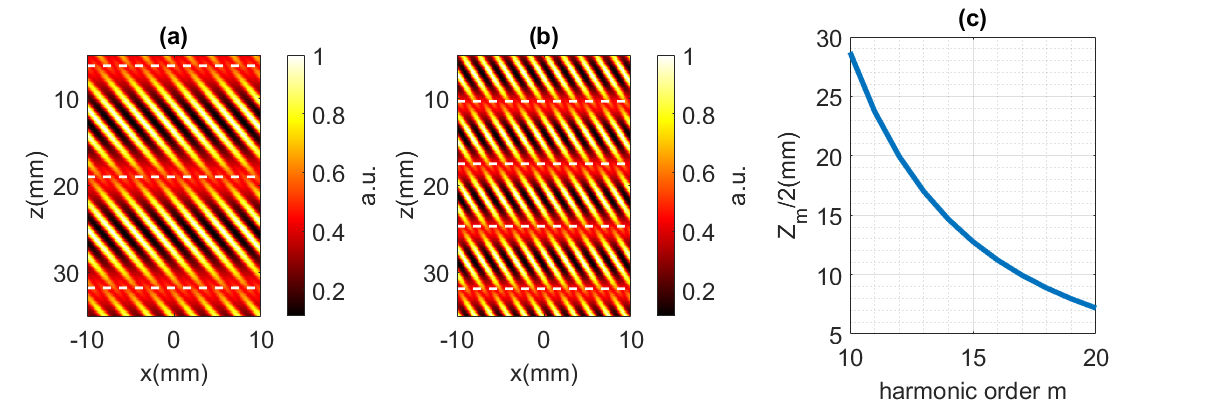}
\caption{Theoretical tagging functions $C_{n=10,m=15}$ (a) and $C_{n=10,m=20}$ (b) as defined in Eq~\ref{eq:CnotCorr}. In both images, horizontal dotted lines are separated by a half-Talbot distance $z_m/2$. (c) Half-Talbot distance $z_m/2$ as a function of harmonic $m$ for $L=38.4~\mathrm{mm}$.}
\label{fig:FigH2}
\end{figure}

\noindent Where $z_m =\frac{ 2\nu_{us}L^2 }{m^2c_s}$ is the Talbot distance. $C_{n,m}$ is plotted for $n=10$ and respectively $m=15$ in Fig~\ref{fig:FigH2}(a) and $m=20$ in Fig~\ref{fig:FigH2}(b). The half-Talbot distance is plotted In Fig~\ref{fig:FigH2}(c) for $10\le m\le 20$. 

%In-phase constructive interference will occurs along the propagation each time the $e^{i\theta_{m}k^2}$ factor reduces to a linear phase shift between all Fourier components. When this condition is met, a replica of the original pattern will appear which may be spatially translated relative to the pattern initially emitted by the probe. Note that for an arbitrarily large number of Fourier components, there are only a limited number of positions where this factorization can occur, namely those such that $\forall k\in \mathbb{Z}$:

%$$
%e^{i\theta_{m}k^2} = e^{i\theta_{m}k}
%$$

%\noindent The resolution of this equation implies to find the $\theta_{m}$ such that $\forall k\in \mathbb{Z}$ we have $\theta_{m}k(k-1) = 0 [2\pi]$. It is quite straightforward that $k(k-1)$ is always an even number ( since either $k$ or $k-1$ is even ), such that we clearly have solutions for:

%$$
%\theta_{m} = p\pi , \ \mathrm{with} \ p \in \mathbb{Z}
%$$

%\noindent which corresponds to the positions:

%\begin{equation}
% z_p = p \frac{\nu_{us}L^2}{c_sm^2} = p \frac{z_m}{2}
%\end{equation}

%\noindent Where $z_m =\frac{ 2\nu_{us}L^2 }{m^2c_s}$ is the Talbot distance. For each of the revival solutions, the phase offset is simply obtained by factorisation of the linear chirp such that we trivial have that any given position $z_p$, the self-imaged periodic pattern is translated by $p \pi$. We there have a $\pi$-offset artefact on the fringes of the tagging function at any $z_p$ distance where $p$ is odd, as illustrated in Fig.~\ref{fig:FigH2}(a).

%\begin{equation}
%C_{n,m}(x,z) = |\sum_{k=-\infty}^{\infty}\left( \sum_{l=-\infty}^{\infty}a^{n,m}_{k,l} e^{j2\pi (\frac{lm}{L})^2\frac{  z c_s}{2\nu_{us}}}e^{-2 j\pi\frac{lm}{L}x}\right) \int_{\tau_e} e^{2j \pi\frac{kn}{T_0}(t-c_s^{-1}z)}\sum_{k'=-\infty}^{\infty}(b^{n}_{k'})^*e^{-2j \pi\frac{k'n}{T_0}t}dt|^2
%\end{equation}
%
%\begin{equation}
%C_{n,m}(x,z) = |\sum_{k=-\infty}^{\infty}\left( \sum_{l=-\infty}^{\infty}a^{n,m}_{k,l} e^{j2\pi (\frac{lm}{L})^2\frac{  z c_s}{2\nu_{us}}}e^{-2 j\pi\frac{lm}{L}x}\right) \sum_{k'=-\infty}^{\infty}(b^{n}_{k'})^*e^{-2j \pi\frac{kn}{T_0}c_s^{-1}z}\int_{\tau_e} e^{2j \pi\frac{(k-k')n}{T_0}t}dt|^2
%\end{equation}

%\begin{equation}
%C_{n,m}(x,z) = \left|\tau_e \sum_{k=-\infty}^{\infty}(b^{n}_{k})^*e^{-2j \pi\frac{kn}{T_0}c_s^{-1}z}\left( \sum_{l=-\infty}^{\infty}a^{n,m}_{k,l} e^{i \theta_m l^2}e^{-2 i\pi\frac{lm}{L}x}\right) \right|^2
%\end{equation}

\noindent Since the imaging method further requires perform linear combinations of measurements using different values of $\phi$, the extra-modulation $\cos(2\pi z/z_m)$ will fundamentally deteriorates the expected Fourier component. We clearly see in Fig~\ref{fig:FigH2}(b) that the first Talbot jump occurs at a distance which decreases with the harmonic orders $m$. This means the degradation in imaging will be more pronounced as objects are positioned away from the probe, which is a problem for a imaging technique intended to image at large depths. We are now going to see a simple way to circumvent this problem.

\subsection{$\pi$-phase-jump corrections}

\begin{figure}[htbp]
\centering\includegraphics[width=\textwidth]{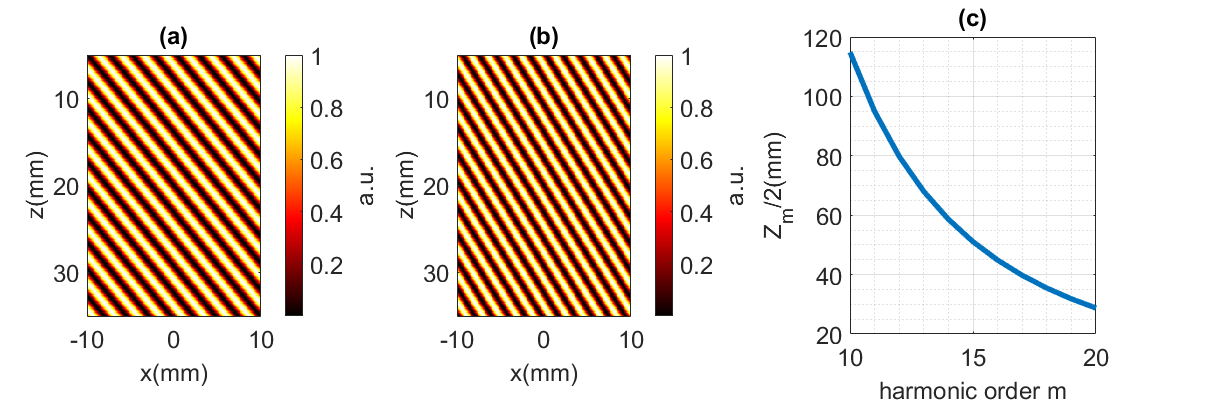}
\caption{Theoretical tagging functions $C_{n=10,m=15}$ (a) and $C_{n=10,m=20}$ (b) as defined in Eq~\ref{eq:Ccorrected}. (c) Half-Talbot distance $z_m/2$ as a function of harmonic $m$ for $L=38.4~\mathrm{mm}$.}
\label{fig:FigH4}
\end{figure}

The appearance of the $\pi$-shifted self-images at every half-Talbot distance from the surface can be removed by imposing an additional phase-jump modulation on the initial periodic pattern. This consists in periodically multiplying the $h_{n}$ by $e^{i\pm\pi}=\pm 1$ at the structuring frequency. The resulting modified function is now $2T_0/n$-periodic and $T_0/n$-antiperiodic such that $a_{2k} = 0 \ \forall k\in \ \mathrm{Z}$. Note that in this case, the Talbot distance is four times higher than in the previous case because of the quadratic dependence of $z_m$ with the structuration period. Although this increase in Talbot distance illustrated in Fig~\ref{fig:FigH4}(b) could already explain an improvement in resolution for a given position below the probe, we see hereafter that this phase correction actually completely suppresses the Talbot artifact.

\begin{figure}[htbp]
\centering\includegraphics[width=0.8\textwidth]{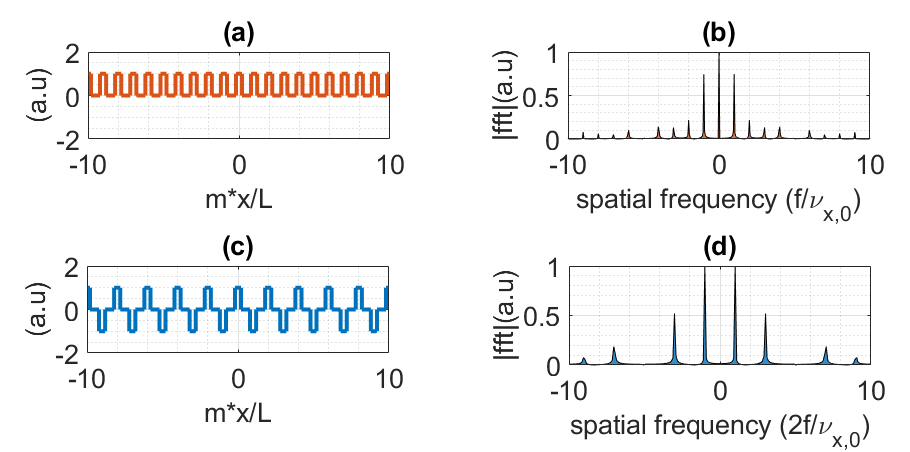} 
\caption{Amplitude modulation of the acoustic emitted field at position $z=0$ for $m=20$ in absence of phase correction (a); with the additional $\pi$-phase modulation (c) and corresponding power spectral densities in both cases (b,d)}
\label{fig:illustration-PhaseJumps}
\end{figure}

Using the previous Eq~\ref{eq:Eq2bis} at $z=0$ and $t=0$, we obtain a similar phase-jump modulation pattern along $x$ as illustrated in Fig~\ref{fig:illustration-PhaseJumps}(b) for $m=20$. Both respective Fourier spectra are shown in (c,d), where we see the annihilation of the even components of the spectrum when the phase correction is applied (d). Likewise, the LO $h^{n,\phi}_{\mathrm{ref}}$ originally defined by Eq~\ref{eq:Href} undergoes the same phase-jump correction so as to remain in phase with $h_n$ during the camera integration time. To assess the effect of this correction on the correlation function, we use the generic expression of the modulation function defined in Eq~\ref{eq:diffHnm}, and simply perform the substitution $T_0 \leftarrow 2T_0$ and $L \leftarrow 2L$, while recalling that we now have $a_{0} =0$, the expression of the correlation function becomes:

\begin{equation}
\label{eq:Ccorrected}
C_{n,m}(x,z) = |a_1e^{i\theta_m/4}\sin(\pi\chi+\phi/2)|^2=|a_1|^2[1-\cos(2\pi\chi_{nm}+\phi)]/2.
\end{equation}

 \noindent Looking at Eq~\ref{eq:Ccorrected}, the interpretation of the Talbot correction becomes quite straightforward: because there is no longer a constant term in the expression, the beating factor $e^{i\theta_m/4}$ can be factorized and is equal to one once we apply the modulus. Another interpretation of this result is to consider that both the negative and positive Talbot carpets cannot be differentiated in the measurement since the modulus of a negative function is always positive. The $\pi$-offset therefore does not appear as a detrimental artifact anymore, as depicted in Fig~\ref{fig:FigH4}, where we plot the tagging function defined by Eq~\ref{eq:Ccorrected}. We can now therefore extract a pure Fourier component as originally described by the method.

 %But there is another point: we can also suppose, without loss of generality, that $a_k^n$ is now an imaginary odd function of $k$ (through a proper choice of $h_n(\cdot)$ as an odd function). If we go back to Eq~\ref{eq:diffHnm}, replacing $L$ by $2L$ and $T_0$ by $2T_0$ in order to account for the doubling of the structuration period, we note that under this condition the modulation function cancels when $\chi$ is an integer. One can therefore trace zero lines of this function on the $(x,z)$ plane, and one could understand that the acoustic energy is constrained to stay located between these lines, forming diagonal bands as pictured in Fig\ref{fig:FigH2}b.

%\begin{equation}
%h_{n,m}(t,x) = \sum_{k=-\infty}^{\infty}\sum_{l=-\infty}^{\infty}a^{n,m}_{k,l} e^{2j \pi\frac{kn}{2T_0}t}e^{-2 j\pi\frac{lm}{2L}x}
%\end{equation}
%
%
%\begin{equation}
%h_{n,m}(t,x+\frac{T_0}{n}) = \sum_{k=-\infty}^{\infty}\sum_{l=-\infty}^{\infty}a^{n,m}_{k,l} e^{2j \pi\frac{kn}{2T_0}t}e^{-2 j\pi\frac{lm}{2L}x}e^{- j\pi l}
%\end{equation}

%\begin{equation}
%\left\{
%    \begin{array}{ll}
%        a^{n,m}_{k,l}(-1)^l = -a^{n,m}_{k,l} \\
%        a^{n,m}_{k,l}(-1)^k = -a^{n,m}_{k,l}
%    \end{array}
%\right.
%\end{equation}

\section{Experimental validation}

\begin{figure}[htbp]
\centering\includegraphics[width=\textwidth]{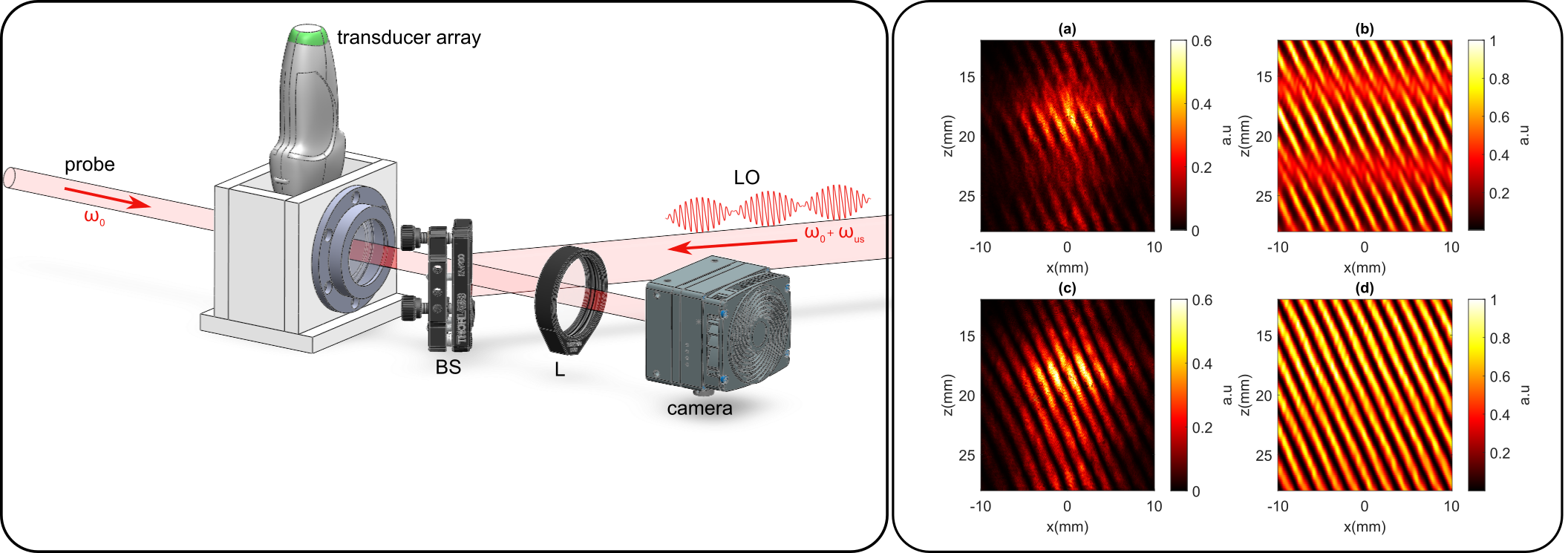} 
\caption{Left caption: experimental setup of ballistic imaging of the tagging function. LO: local oscillator, BS: beam splitter, L: lens with a focal of $100~\mathrm{mm}$.  Right caption: comparison of the tagging function for $n=10$ and $m=20$. (a) Experimental measurement without phase-jump correction; (b) Simulation without phase-jump correction; (c) Experimental measurement with phase-jump correction;(d) Simulation with phase-jump correction.}
\label{fig:comparison}
\end{figure}

We propose the simple experiment scheme depicted in Figure~\ref{fig:comparison} to image the tagging functions and test our Talbot correction method. A CW laser beam centered at $780~\mathrm{nm}$ of $\sim 4~\mathrm{\mu W}$ propagates through a water tank sealed on each side by two optically polished silica windows. A linear transducer array, composed of 192 elements with a $200~\mathrm{\mu m}$ pitch (SL10-2 supersonic), is encoded to send long ultrasonic bursts centered at $\nu_{us}=3~\mathrm{MHz }$ with a given modulation $h_{n,m}$. The fundamental frequencies are set to $\nu_{x0} = 26.04~\mathrm{m^{-1}}$ and $\nu_{z0} = 32.46~\mathrm{m^{-1}}$, as in~\cite{dutheil2021fourier}.
 A $f=100~\mathrm{mm}$ lens is positioned in the beam path to conjugate the ultrasonic imaging plane with the sensor of a digital camera (Ximea Xib-64, 300FPS). The latter beam is overlapped on the camera sensor with a LO beam centered at the tagged photons frequency and modulated in amplitude by $h^{n,\phi}_{\mathrm{ref}}$ by means of an acousto-optic modulator (not represented here), so as to perform an off-axis holographic detection. The camera exposure time is set to $\tau_{c} = 200~\mathrm{\mu s}$ for all acquired frames, at a $300~\mathrm{Hz}$ repetition rate. Each acquired image is filtered in the Fourier domain by isolating the first off-axis order resulting from the constructive interference of the tagged photons with the LO. The filtered image is then translated back into real space so that only tagged photons are visible. This allows to visualize the tagging function for a given value $(n,m)$. 

We simulate the acoustic wave propagation in water using FieldII open-source simulation software~\cite{jensen1991model} already used and described in a previous work~\cite{bocoum2019structured}. As a result, the correlation with the modulated reference beam can be computed, thereby providing a simulated image of the tagging functions for each harmonic $(n,m)$. A comparison between the experimentally measured and the simulated tagging function is depicted on the right caption of Fig~\ref{fig:comparison} for $n=10$ and $m=20$ for $10\le z(~\mathrm{mm}) \le 30$ and $-10\le x(\mathrm{mm}) \le 10$. In both Fig~\ref{fig:comparison} (a) and (b), where no correction is applied to the US pulse, we clearly distinguish $\pi$-Talbot jumps precisely located at $z_T/2$, as expected. By performing both the same measurement and simulation while imposing the phase jump correction on the US as previously described, we obtain Fig~\ref{fig:comparison} (c) and (d) where we observe a complete suppression of this Talbot artifact. The same observation holds for all values of $10\le m\le 20$ we have measured. Both our simulation and experimental observations are well in agreement with our previous model, thereby demonstrating the effectiveness of our correction method.

%\begin{figure}[htbp]
%\centering\includegraphics[width=12cm]{../images/NbXNbZ20_imageInclusions.png}
%\caption{Reconstruction of a simulated acquisition with stand structrured waves (a) and with addition phase-jump modulation (b)}
%\end{figure}

\begin{figure}[htbp]
\centering\includegraphics[width=\textwidth]{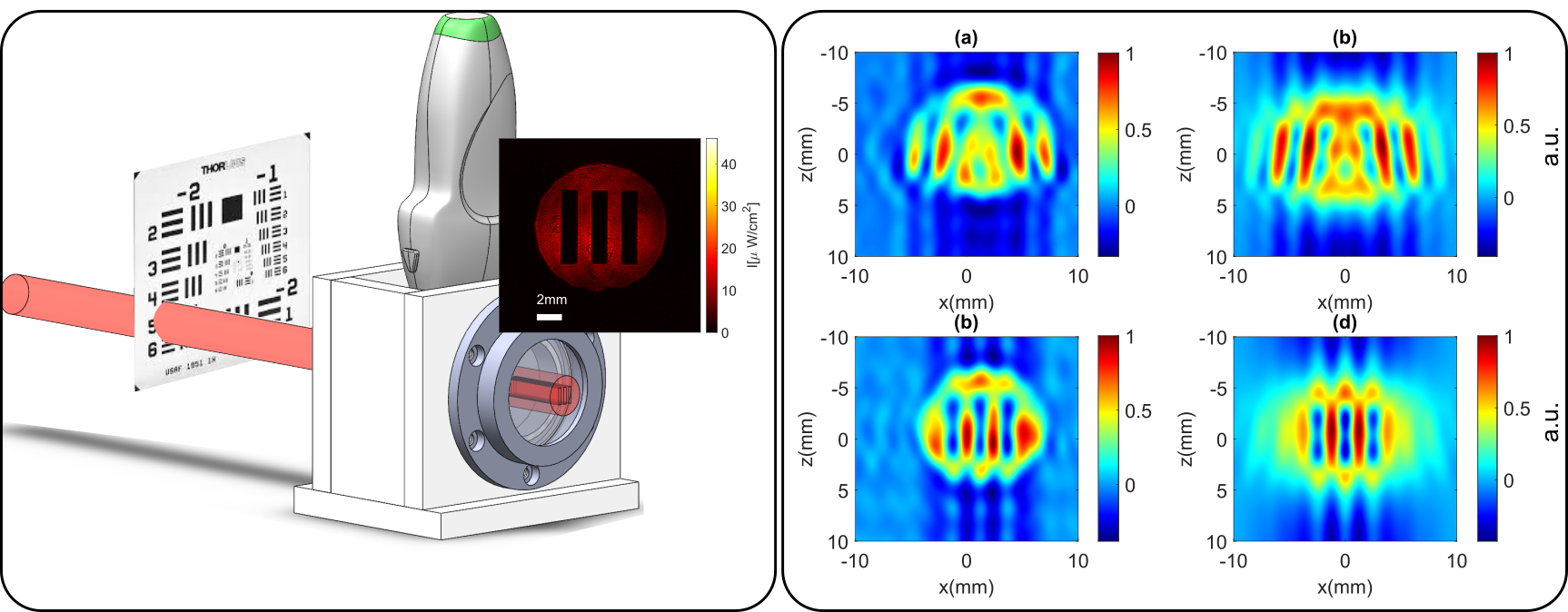}
\caption{Left caption: experimental setup of the shadowgraphed USAF resolution target positioned on the vertical line with Group number -2 Element 5, corresponding to a linewidth of $1.26~\mathrm{mm}$ in the US imaging plane. Right caption: reconstructed images without Talbot correction (a) and comparison with the simulated reconstruction for a beam waist of $8~\mathrm{mm}$ (b). Experimental (b) and simulated (d) in the exact same conditions when implementing Talbot phase-jump corrections.}
\label{fig:image}
\end{figure}

\noindent We first propose to test a full image reconstruction by inserting a resolution target (Thorlabs USAF) on vertical lines corresponding to the Group number -2 Element 5 in the beam path of the probe before it enters in the water tank. This corresponds to a line width of $1.26~\mathrm{mm}$. This setup allows to create a test object by shadowgraphy as the resolution target is imprinted in the imaging plane defined by the position of the US transducer, as depicted in the left caption of Fig~\ref{fig:image}. The reconstructed results with (b) and without (a) Talbot correction are shown for an acquisition using $1\le n\le 20$, $-20 \le m \le 20$ and four phases for each couple $(n,m)$. This is a total of $4\times20\times21 = 1680$ frames to acquire an image, leading to an acquisition time of $5.6~\mathrm{s}$. The corresponding simulated reconstructions are shown for comparison with (d) and without (b). The excellent qualitative agreement between the experimental and simulated reconstructions are clear evidence that the Talbot effect causes strong deformation of the image, since the target appears to be made of four rather than three lines. When the correction is applied, we observe a significant improvement of the image and the resolution target is well retrieved. 

\begin{figure}[htbp]
\centering\includegraphics[width=12cm]{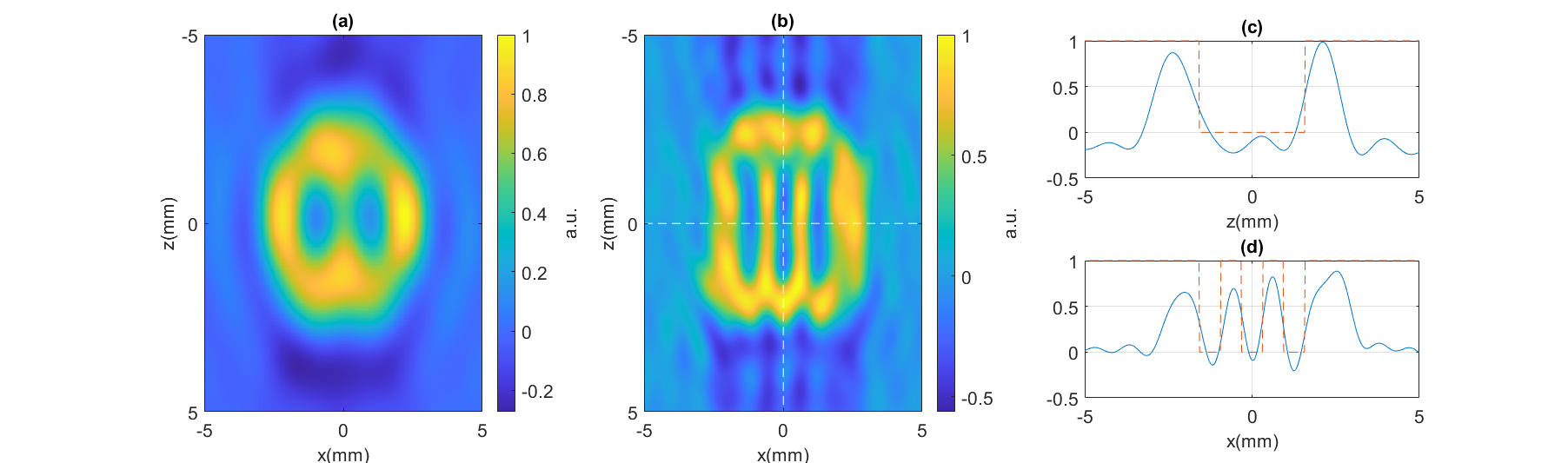}
\caption{Image reconstruction for a resolution target with a $0.63~\mathrm{mm}$ linewidth for $1\le n\le 20$ and $-20 \le m \le 20$ using (a) $\nu_{x0} = 26.04~\mathrm{m^{-1}}$ and (b)$\nu_{z0} = 32.46~\mathrm{m^{-1}}$ and $\nu_{x0} = 52.08~\mathrm{m^{-1}}$ and $\nu_{z0} = 64.92~\mathrm{m^{-1}}$. (c) Vertical line profile $x=0$ overlapped with the theoretical target represented by a dotted line (d) horizontal line profile $z=0$ overlapped with the theoretical target represented by a dotted line}
\label{fig:imageZoom}
\end{figure}

Finally, to test the limit in resolution of the method without increasing the number of components, we move the calibration target to the vertical lines with Group number -1 Element 5, which corresponds to a linewidth of $0.63~\mathrm{mm}$, that is to say twice smaller than the one used for Fig~\ref{fig:image}. Using the same probing parameters as previously, we obtain the image  Fig~\ref{fig:imageZoom}(a), which clearly shows the three vertical lines cannot be resolved. Therefore, we reiterated the experiment by doubling the fundamental frequencies used to $\nu_{x0} = 52.08~\mathrm{m^{-1}}$ and $\nu_{z0} = 64.92~\mathrm{m^{-1}}$ for the same probing values $1\le n\le 20$ and $-20 \le m \le 20$. Doing so, we now have a resolution of $480~\mathrm{\mu m}$ along $x$ and $385~\mathrm{\mu m}$ along $z$ that compares with the ultrasonic central wavelength $\lambda_{us}=513~\mathrm{\mu m}$. The experimental reconstruction is shown in Fig~\ref{fig:imageZoom}(b) where we clearly see the three vertical lines now clearly resolved.

\section{Conclusion}

We have demonstrated the possibility to reach near-diffraction-limited imaging resolution using plane wave structured insonification in the context of acousto-optic imaging. So far, a limit to improving the imaging resolution along direction $x$ was the appearance of Talbot artifacts as we increased the structuring frequency along the transducer array. We have shown how imposing a $\pi$-phase jump on the periodic pattern could remove this artifact without affecting the tagging function resulting from the initial US modulation. Doing so, we have improved by a factor close to ten the image resolution of our technique, reaching near-diffraction resolution. This result is an important new step towards the implementation of AO imaging based on holographic detection compatible with \textit{in-vivo} decorrelation time scales. The method will be implement on living mice in a near future.

\begin{backmatter}

\bmsection{Funding} French LABEX WIFI, ANR-21-CE42-0014.

\bmsection{Disclosures} The authors declare no conflicts of interest.

\bmsection{Data availability} Data underlying the results presented in this paper are not publicly available at this time but may be obtained from the authors upon reasonable request.
\end{backmatter}

%%%%%%%%%% If using BibTeX:
\bibliography{sample}

\begin{thebibliography}{10}
\newcommand{\enquote}[1]{``#1''}

\bibitem{bocoum2019structured}
M.~Bocoum, J.-L. Gennisson, J.-B. Laudereau, \emph{et~al.}, \enquote{Structured
  ultrasound-modulated optical tomography,} {\protect\JournalTitle{Applied
  optics}} \textbf{58}, 1933--1940 (2019).

\bibitem{bocoum2020reconstruction}
M.~Bocoum, J.-L. Gennisson, A.~A. Grabar, \emph{et~al.},
  \enquote{Reconstruction of bi-dimensional images in fourier-transform
  acousto-optic imaging,} {\protect\JournalTitle{Optics Letters}} \textbf{45},
  4855--4858 (2020).

\bibitem{dutheil2021fourier}
L.~Dutheil, M.~Bocoum, M.~Fink, \emph{et~al.}, \enquote{Fourier transform
  acousto-optic imaging with off-axis holographic detection,}
  {\protect\JournalTitle{Applied optics}} \textbf{60}, 7107--7112 (2021).

\bibitem{qureshi2017vivo}
M.~M. Qureshi, J.~Brake, H.-J. Jeon, \emph{et~al.}, \enquote{In vivo study of
  optical speckle decorrelation time across depths in the mouse brain,}
  {\protect\JournalTitle{Biomedical optics express}} \textbf{8}, 4855--4864
  (2017).

\bibitem{liu2015optical}
Y.~Liu, P.~Lai, C.~Ma, \emph{et~al.}, \enquote{Optical focusing deep inside
  dynamic scattering media with near-infrared time-reversed ultrasonically
  encoded (true) light,} {\protect\JournalTitle{Nature communications}}
  \textbf{6}, 5904 (2015).

\bibitem{cheng2023high}
Z.~Cheng, C.~Li, A.~Khadria, \emph{et~al.}, \enquote{High-gain and high-speed
  wavefront shaping through scattering media,} {\protect\JournalTitle{Nature
  Photonics}} pp. 1--7 (2023).

\bibitem{talbot1836lxxvi}
H.~F. Talbot, \enquote{Lxxvi. facts relating to optical science. no. iv,}
  {\protect\JournalTitle{The London, Edinburgh, and Dublin Philosophical
  Magazine and Journal of Science}} \textbf{9}, 401--407 (1836).

\bibitem{leger1989lateral}
J.~R. Leger, \enquote{Lateral mode control of an {A}l{G}a{A}s laser array in a
  talbot cavity,} {\protect\JournalTitle{Applied physics letters}} \textbf{55},
  334--336 (1989).

\bibitem{leger1988coherent}
J.~R. Leger, M.~L. Scott, and W.~B. Veldkamp, \enquote{Coherent addition of
  {A}l{G}a{A}s lasers using microlenses and diffractive coupling,}
  {\protect\JournalTitle{Applied physics letters}} \textbf{52}, 1771--1773
  (1988).

\bibitem{mehuys1991modal}
D.~Mehuys, W.~Streifer, R.~G. Waarts, and D.~F. Welch, \enquote{Modal analysis
  of linear talbot-cavity semiconductor lasers,} {\protect\JournalTitle{Optics
  letters}} \textbf{16}, 823--825 (1991).

\bibitem{chowdhury2018structured}
S.~Chowdhury, J.~Chen, and J.~A. Izatt, \enquote{Structured illumination
  fluorescence microscopy using talbot self-imaging effect for high-throughput
  visualization,} {\protect\JournalTitle{arXiv preprint arXiv:1801.03540}}
  (2018).

\bibitem{stuerzebecher2010advanced}
L.~Stuerzebecher, T.~Harzendorf, U.~Vogler, \emph{et~al.}, \enquote{Advanced
  mask aligner lithography: fabrication of periodic patterns using pinhole
  array mask and talbot effect,} {\protect\JournalTitle{Optics express}}
  \textbf{18}, 19485--19494 (2010).

\bibitem{wang2016sub}
L.~Wang, F.~Clube, C.~Dais, \emph{et~al.}, \enquote{Sub-wavelength printing in
  the deep ultra-violet region using displacement talbot lithography,}
  {\protect\JournalTitle{Microelectronic Engineering}} \textbf{161}, 104--108
  (2016).

\bibitem{rayleigh1881xxv}
L.~Rayleigh, \enquote{Xxv. on copying diffraction-gratings, and on some
  phenomena connected therewith,} {\protect\JournalTitle{The London, Edinburgh,
  and Dublin Philosophical Magazine and Journal of Science}} \textbf{11},
  196--205 (1881).

\bibitem{dennis2007plasmon}
M.~R. Dennis, N.~I. Zheludev, and F.~J.~G. De~Abajo, \enquote{The plasmon
  talbot effect,} {\protect\JournalTitle{Optics Express}} \textbf{15},
  9692--9700 (2007).

\bibitem{chapman1995near}
M.~S. Chapman, C.~R. Ekstrom, T.~D. Hammond, \emph{et~al.}, \enquote{Near-field
  imaging of atom diffraction gratings: The atomic talbot effect,}
  {\protect\JournalTitle{Physical Review A}} \textbf{51}, R14 (1995).

\bibitem{clauser1994talbot}
J.~F. Clauser and S.~Li, \enquote{Talbot-vonlau atom interferometry with cold
  slow potassium,} {\protect\JournalTitle{Physical Review A}} \textbf{49},
  R2213 (1994).

\bibitem{deng1999temporal}
L.~Deng, E.~W. Hagley, J.~Denschlag, \emph{et~al.}, \enquote{Temporal,
  matter-wave-dispersion talbot effect,} {\protect\JournalTitle{Physical Review
  Letters}} \textbf{83}, 5407 (1999).

\bibitem{morozov2017observation}
A.~N. Morozov, M.~P. Krikunova, B.~Skuibin, and E.~V. Smirnov,
  \enquote{Observation of the talbot effect for ultrasonic waves,}
  {\protect\JournalTitle{JETP Letters}} \textbf{106}, 23--25 (2017).

\bibitem{candelas2019observation}
P.~Candelas, J.~M. Fuster, S.~P{\'e}rez-L{\'o}pez, \emph{et~al.},
  \enquote{Observation of ultrasonic talbot effect in perforated plates,}
  {\protect\JournalTitle{Ultrasonics}} \textbf{94}, 281--284 (2019).

\bibitem{latimer1992talbot}
P.~Latimer and R.~F. Crouse, \enquote{Talbot effect reinterpreted,}
  {\protect\JournalTitle{Applied optics}} \textbf{31}, 80--89 (1992).

\bibitem{berry2001quantum}
M.~Berry, I.~Marzoli, and W.~Schleich, \enquote{Quantum carpets, carpets of
  light,} {\protect\JournalTitle{Physics World}} \textbf{14}, 39 (2001).

\bibitem{wen2013talbot}
J.~Wen, Y.~Zhang, and M.~Xiao, \enquote{The talbot effect: recent advances in
  classical optics, nonlinear optics, and quantum optics,}
  {\protect\JournalTitle{Advances in optics and photonics}} \textbf{5}, 83--130
  (2013).

\bibitem{pelka2018prime}
K.~Pelka, J.~Graf, T.~Mehringer, and J.~von Zanthier, \enquote{Prime number
  decomposition using the talbot effect,} {\protect\JournalTitle{Optics
  Express}} \textbf{26}, 15009--15014 (2018).

\bibitem{de2013theory}
H.~G. De~Chatellus, E.~Lacot, W.~Glastre, \emph{et~al.}, \enquote{Theory of
  talbot lasers,} {\protect\JournalTitle{Physical Review A}} \textbf{88},
  033828 (2013).

\bibitem{romero2019arbitrary}
L.~Romero~Cort{\'e}s, R.~Maram, H.~Guillet~de Chatellus, and J.~Aza{\~n}a,
  \enquote{Arbitrary energy-preserving control of optical pulse trains and
  frequency combs through generalized talbot effects,}
  {\protect\JournalTitle{Laser \& photonics reviews}} \textbf{13}, 1900176
  (2019).

\bibitem{jensen1991model}
J.~A. Jensen, \enquote{A model for the propagation and scattering of ultrasound
  in tissue,} {\protect\JournalTitle{The Journal of the Acoustical Society of
  America}} \textbf{89}, 182--190 (1991).

\end{thebibliography}

%%%%%%%%%% If preparing manually:
% \begin{thebibliography}{1}
% \newcommand{\enquote}[1]{``#1''}

% \bibitem{Zhang:14}
% Y.~Zhang, S.~Qiao, L.~Sun, Q.~W. Shi, W.~Huang, L.~Li, and Z.~Yang,
%   \enquote{Photoinduced active terahertz metamaterials with nanostructured
%   vanadium dioxide film deposited by sol-gel method,}
%   {\protect\JournalTitle{Optics Express}} \textbf{22}, 11070--11078 (2014).

% \bibitem{Optica}
% {Optica}, \enquote{{Optica Publishing Group},}
%   \url{http://www.opg.optica.org}.

% \bibitem{FORSTER2007}
% P.~Forster, V.~Ramaswamy, P.~Artaxo, T.~Bernsten, R.~Betts, D.~Fahey,
%   J.~Haywood, J.~Lean, D.~Lowe, G.~Myhre, J.~Nganga, R.~Prinn, G.~Raga,
%   M.~Schulz, and R.~V. Dorland, \enquote{Changes in atmospheric consituents and
%   in radiative forcing,} in \enquote{Climate Change 2007: The Physical Science
%   Basis. Contribution of Working Group 1 to the Fourth assesment report of
%   Intergovernmental Panel on Climate Change,}  S.~Solomon, D.~Qin, M.~Manning,
%   Z.~Chen, M.~Marquis, K.~B. Averyt, M.~Tignor, and H.~L. Miler, eds.
%   (Cambridge University Press, 2007).

% \end{thebibliography}

\end{document}